\documentclass[aps,preprint,tightenlines,groupedaddress,nofootinbib,byrevtex,showpacs]{revtex4}
\usepackage{graphicx,comment}

\usepackage{amssymb,latexsym}
\usepackage{amsmath,amsbsy}

\begin{document}

\unitlength=1mm

\def\a{{\alpha}}
\def\b{{\beta}}
\def\d{{\delta}}
\def\D{{\Delta}}
\def\e{{\epsilon}}
\def\g{{\gamma}}
\def\G{{\Gamma}}
\def\k{{\kappa}}
\def\l{{\lambda}}
\def\L{{\Lambda}}
\def\m{{\mu}}
\def\n{{\nu}}
\def\o{{\omega}}
\def\O{{\Omega}}
\def\S{{\Sigma}}
\def\s{{\sigma}}
\def\th{{\theta}}

\def\ol#1{{\overline{#1}}}

\def\Dslash{D\hskip-0.65em /}
\def\dslash{{\partial\hskip-0.5em /}}
\def\vslash{{\rlap \slash v}}

\def\CPT{{$\chi$PT}}
\def\QCPT{{Q$\chi$PT}}
\def\PQCPT{{PQ$\chi$PT}}
\def\tr{\text{tr}}
\def\str{\text{str}}
\def\diag{\text{diag}}
\def\order{{\mathcal O}}
\def\vit{{\it v}}
\def\vD{\vit\cdot D}
\def\am{\alpha_M}
\def\bm{\beta_M}
\def\gm{\gamma_M}
\def\smb{\sigma_M}
\def\smt{\overline{\sigma}_M}
\def\tb{{\tilde b}}

\def\cS{{\mathcal S}}
\def\cC{{\mathcal C}}
\def\cB{{\mathcal B}}
\def\cT{{\mathcal T}}
\def\cQ{{\mathcal Q}}
\def\cL{{\mathcal L}}
\def\cO{{\mathcal O}}
\def\cA{{\mathcal A}}
\def\cH{{\mathcal H}}
\def\cF{{\mathcal F}}
\def\cG{{\mathcal G}}
\def\cE{{\mathcal E}}
\def\cJ{{\mathcal J}}
\def\cK{{\mathcal K}}
\def\cM{{\mathcal{M}_+}}

\def\Bbar{\overline{B}}
\def\Tbar{\overline{T}}
\def\cBbar{\overline{\cal B}}
\def\cTbar{\overline{\cal T}}
\def\cA{\mathcal A}
\def\pq{(PQ)}

\def\eqref#1{{(\ref{#1})}}

%\preprint{DUKE-TH-04-XXX}
 
\title{Baryon Masses in Partially Quenched Heavy Hadron Chiral Perturbation
  Theory}
\author{ Brian C.~Tiburzi}
\email[]{bctiburz@phy.duke.edu}
\affiliation{Department of Physics\\
Duke University\\
P.O.~Box 90305\\
Durham, NC 27708-0305}

\date{\today}

\begin{abstract} 
The masses of baryons containing a heavy quark 
are calculated to next-to-leading order in partially quenched 
heavy hadron chiral perturbation theory. Calculations 
are performed for three light flavors in the isospin limit
and additionally for two light non-degenerate flavors. 
The results presented are necessary for extrapolating 
lattice QCD and partially quenched lattice QCD calculations 
of the heavy hadron masses.
\end{abstract}

\maketitle

\section{Introduction}

Understanding properties of hadrons directly from QCD represents a considerable on-going challenge. 
A feature of strong interaction dynamics is the confinement of quarks and gluons 
into color-neutral hadronic states, and their spectrum is a fundamental observable
of the theory. Calculation of hadronic masses will be a milepost in strong interaction physics.
For baryons containing $b$ or $c$ quarks, the underlying dynamics encompasses the interplay
of heavy-quark physics and light-quark physics, which are opposite limits of QCD.

Lattice QCD provides a first principles tool for numerical determination of QCD observables, for example, 
the mass spectrum of baryons containing a heavy quark. Calculations of singly heavy baryon masses 
have been pursued in quenched QCD (QQCD) 
QCD~\cite{Alexandrou:1994dm,Bowler:1996ws,AliKhan:1999yb,Woloshyn:2000fe,Lewis:2001iz,Mathur:2002ce} using 
heavy quark effective actions.
Up until now~\cite{Toussaint:2004cj}, partially quenched QCD (PQQCD) calculations in 
the baryonic sector have been very limited.  
The necessity for quenched and partially quenched variations approximations in lattice QCD arises from computational difficulties 
inherent to calculating the fermionic determinant. In QQCD, the time-costly determinant is replaced by a constant,
whereas in PQQCD the determinant is calculated by using larger quark masses 
for the quarks not coupled to external sources. In PQQCD, sea quark contributions are thus retained;
while in QQCD, the sea quarks have been unphysically integrated out. Due to restrictive computational time, 
the valence quark masses, too, cannot be set to their physical values and one must extrapolate down from the values 
used on the lattice.

To perform rigorous extrapolations in the quark mass, one must understand how QCD observables
behave as the quark masses vary. For low-energy QCD, there exists a model-independent effective theory, 
chiral perturbation theory (\CPT), that is written in terms of the pseudo-Goldstone bosons appearing from chiral 
symmetry breaking~\cite{Gasser:1983yg,Gasser:1985gg}. 
While these bosons are not massless, they remain light compared to the scale of chiral symmetry breaking
and hence dominate low-energy hadronic interactions.
Because the light quark masses explicitly break chiral symmetry, they appear as parameters of \CPT\ and enable
the determination of the quark mass dependence of QCD observables by matching onto the effective theory. 
Without external input (e.g.~experimental data or lattice QCD data), 
the effective theory generally lacks complete predictive power as symmetries  
constrain the types of operators in the Lagrangian, not the values of their coefficients.
While providing a guide to extrapolate lattice QCD data, \CPT\  
also benefits from the determination of its low-energy constants (LECs) from these data. 
Furthermore heavy quark symmetry can be combined with chiral symmetry to describe
hadrons formed from both heavy and light quarks~\cite{Burdman:1992gh,Wise:1992hn,Yan:1992gz}.

For QQCD lattice simulations, such as those for singly heavy baryons, 
quenched chiral perturbation theory (\QCPT) has been developed for the 
extrapolation in valence quark 
masses~\cite{Morel:1987xk,Sharpe:1992ft,Bernard:1992mk,Sharpe:1996qp,Labrenz:1996jy}. There is, however, no
general connection of QQCD observables to QCD because QQCD lacks an axial anomaly. 
As a manifestation of this fact, \QCPT\ contains operators that do not have analogues in \CPT. 
Furthermore, the LECs of the operators that have analogues
are numerically unrelated to their counterparts in \CPT. The quenched approximation
can be remedied by performing partially quenched lattice simulations. 
For such simulations, partially quenched chiral perturbation theory \PQCPT\ has been 
constructed to perform the extrapolation in both sea and valence quark 
masses~\cite{Bernard:1994sv,Sharpe:1997by,Golterman:1998st,Sharpe:2000bc,Sharpe:2001fh,Sharpe:2003vy}. 
Unlike QQCD, PQQCD retains an axial anomaly and the flavor singlet field can be integrated out. 
Consequently the LECs of \CPT\ appear in \PQCPT, and hence there is a means to determine \CPT\ 
parameters using PQQCD lattice simulations. Much work has resulted from this possibility, e.g., in 
the baryonic sector 
see~\cite{Chen:2001yi,Beane:2002vq,Chen:2002bz,Arndt:2003vx,Arndt:2003ww,Arndt:2003we,Arndt:2003vd,Walker-Loud:2004hf,Tiburzi:2004rh}.

We anticipate the inclusion of effects from dynamical light quarks into the singly heavy baryon sector,
and in this work we determine the masses of baryons containing a heavy quark at next-to leading
order (NLO) in partially quenched heavy hadron \CPT. Throughout we work in the heavy-quark limit.
Higher-order corrections can be considered if one wishes to make contact with simulations that go 
beyond the leading-order heavy quark effective action. 
We determine the masses for three-flavor and two-flavor partially quenched simulations
and use the isospin limit in the former. Such expressions allow one to perform
the required extrapolation in light-quark masses. 
Additionally the LECs appearing at NLO in heavy hadron \CPT\ can be determined. 
We spell out explicitly the LECs that appear in the mass splitting of the 
$s_\ell = 0$ and $s_\ell = 1$ baryons from chiral effects to this order,
and investigate the chiral contributions to the mass relation among the sextet baryons.

The organization of the paper is as follows. First in Sec.~\ref{pqhhcpt}, we include 
baryons containing a single heavy quark into \PQCPT. We write down the terms of the $SU(6|3)$ heavy hadron chiral Lagrangian 
that are relevant for the determination of baryon masses at NLO. The masses of the $s_\ell = 0$ 
baryons are calculated in Sec.~\ref{triplet}, while those of the $s_\ell = 1$ baryons are calculated 
in Sec.~\ref{six}. The corresponding results for $SU(4|2)$ with non-degenerate light quarks 
are presented in Appendix~\ref{pqsutwo}. For reference, we give 
the masses in $SU(3)$ and $SU(2)$ \CPT\ in Appendix~\ref{cpt}.  
The summary, Sec.~\ref{summy}, highlights the goal of determining 
the mass splittings between baryons containing a single heavy quark.

\section{\PQCPT\ for heavy hadrons} \label{pqhhcpt}

In this section, we formulate \PQCPT\ for heavy hadrons in $SU(6|3)$. 
We begin by reviewing PQQCD. Next we review the pseudo-Goldstone
mesons of \PQCPT, and finally include the heavy hadrons
into this partially quenched theory.

\subsection{PQQCD} \label{subpqqcd}

In PQQCD, the Lagrangian in the light-quark sector is
\begin{equation}
\mathfrak{L} = \sum_{j,k=1}^9 \ol{q}_j \left(
  i\Dslash - m_q \right)_{jk} q_k
.\label{eq:pqqcdlag}
\end{equation}
This differs from the usual three light-flavor Lagrangian of QCD by the
inclusion of six extra quarks; three bosonic ghost quarks, ($\tilde u,
\tilde d, \tilde s $), and three fermionic sea quarks, ($j, l, r$), in addition
to the light physical quarks ($u, d, s$).  The nine quark fields transform in the 
fundamental representation of the graded $SU(6|3)$ group~\cite{BahaBalantekin:1980pp,BahaBalantekin:1981qy}.  
They have been placed in the nine-component vector
\begin{equation}
q = (u, d, s, j, l, r, \tilde{u}, \tilde{d}, \tilde{s})^{\text{T}}
.\end{equation}
These quark fields obey graded equal-time commutation relations
\begin{equation}
q^\a_i(\mathbf x) q^{\beta \dagger}_j(\mathbf y) -
(-1)^{\eta_i \eta_j} q^{\b \dagger}_j(\mathbf y) q^\a_i(\mathbf x) =
\d^{\a \b} \d_{ij} \d^3 (\mathbf {x-y}),
\label{eq:qetcr}\end{equation}
where $\a, \beta$ and $i,j$ are spin and flavor indices, respectively.
The vanishing graded equal-time commutation relations are written analogously.
Above we have made use of grading factors
\begin{equation}
   \eta_k
   = \left\{ 
       \begin{array}{cl}
         1 & \text{for } k=1,2,3,4,5,6 \\
         0 & \text{for } k=7,8,9
       \end{array} 
     \right.
,\end{equation}
that take into account the different statistics of the PQQCD quark fields.  
When $m_u = m_d$, the quark mass matrix of $SU(6|3)$ is given by
\begin{equation}
m_q = \diag(m_u, m_u, m_s, m_j, m_j, m_r, m_u, m_u, m_s).
\end{equation}
Notice that each valence quark mass is identically equal to its corresponding ghost quark's mass. 
Thus there is an exact cancellation in the path integral between their
respective determinants. The only contributions to observables from 
disconnected quark loops come from the sea sector of PQQCD. Thus one can keep valence 
and sea contributions separate and additionally vary the masses of the valence and 
sea sectors independently. In the limit $m_j \rightarrow m_u$ and $m_r \rightarrow m_s$, 
QCD is recovered.

\subsection{Mesons of \PQCPT} \label{mesonpqcpt}

In the limit that the light quarks become massless, 
the theory corresponding to the Lagrangian in
Eq. (\ref{eq:pqqcdlag}) has a graded symmetry of the form
$SU(6|3)_L \otimes SU(6|3)_R \otimes U(1)_V$, which is
assumed to be spontaneously broken down to 
a remnant $SU(6|3)_V \otimes U(1)_V$ symmetry in analogy with QCD.  
One can build an effective low-energy theory of PQQCD by
perturbing about the physical vacuum state. 
This theory is PQ$\chi$PT, and the emerging pseudo-Goldstone mesons of this theory 
have dynamics described at leading order in the chiral expansion by the Lagrangian
\begin{equation}
  \mathfrak{L} =
    \frac{f^2}{8} \str \left(
      \partial^\mu \S^\dagger \partial_\mu \S \right)
      + \l  \, \str \left( m_q \S^\dagger + m_q^\dagger \S \right)
           +\a_\Phi \partial^\mu \Phi_0 \partial_\mu \Phi_0
           - m_0^2 \Phi_0^2,
\label{eq:pqbosons}
\end{equation}
where the field 
\begin{equation}
  \S = \exp \left( \frac{2 i \Phi}{f} \right) = \xi^2,
\end{equation}
and the meson fields appear in the $SU(6|3)$ matrix
\begin{equation}
    \Phi =
    \begin{pmatrix}
      M & \chi^\dagger\\
      \chi & \tilde M\\
    \end{pmatrix}. \label{eq:mesonmatrix}
\end{equation}
The quantities $\a_\Phi$ and $m_0$ are non-vanishing in the chiral limit.  
The $M$ and $\tilde M$ matrices contain bosonic mesons (with quantum numbers of $q \bar{q}$ pairs and 
$\tilde{q} \bar{\tilde{q}}$ pairs, respectively), while the $\chi$ and $\chi^\dagger$
matrices contain fermionic mesons (with quantum numbers of $\tilde q \bar{q}$
pairs and $q \bar{\tilde{q}}$ pairs, respectively).
The upper $3 \times 3$ block of the matrix $M$ contains the familiar
pions, kaons, and eta, while the remaining components consist of mesons formed
from one or two sea quarks, see e.g.~\cite{Chen:2001yi}.
The operation $\str( )$ in Eq. (\ref{eq:pqbosons}) is a supertrace over flavor indices.  
Expanding the Lagrangian Eq.~\eqref{eq:pqbosons} to leading order,
one finds that mesons with quark content $qq'$ are canonically normalized
when their masses are given by
\begin{equation}
m^2_{qq'} = \frac{4 \lambda}{f^2} (m_q + m_{q'}) 
\label{eq:pqmesonmass}.
\end{equation}

The flavor singlet field that appears above is defined to be 
$\Phi_0 = {\rm str}( \Phi ) / {\sqrt 6}$.  
As PQQCD has a strong axial anomaly $U(1)_A$, 
the mass of the singlet field $m_0$ can be taken to be
on the order of the chiral symmetry breaking scale, and thus $\Phi_0$ 
in Eq.~\eqref{eq:pqbosons} is integrated out of the theory.  
The resulting $\eta$ two-point correlation functions, however, deviate from their familiar form in \CPT.
For $a,b = u,d,s,j,l,r,\tilde u,\tilde d,\tilde s$, 
the leading-order $\eta_a \eta_b$ propagator with $2+1$ sea-quarks is given by
\begin{equation}
{\cal G}_{\eta_a \eta_b} =
        \frac{i \epsilon_a \delta_{ab}}{q^2 - m^2_{\eta_a} +i\epsilon}
        - \frac{i}{3} \frac{\epsilon_a \epsilon_b \left(q^2 - m^2_{jj}
            \right) \left( q^2 - m^2_{rr} \right)}
            {\left(q^2 - m^2_{\eta_a} +i\epsilon \right)
             \left(q^2 - m^2_{\eta_b} +i\epsilon \right)
             \left(q^2 - m^2_X +i\epsilon \right)}\, ,
\end{equation}
where
\begin{equation}
\epsilon_a = (-1)^{1+\eta_a}
.\end{equation}
The mass $m_{\eta_x}$ is the mass of a quark-basis meson composed of $x$ and $\ol x$
quarks, while the mass $m_X$ is defined as $m_X^2 =
\frac{1}{3}\left(m^2_{jj} + 2 m^2_{rr}\right)$.  The flavor neutral  
propagator can be conveniently rewritten as
\begin{equation}
{\cal G}_{\eta_a \eta_b} =
         \e_a \d_{ab} P_a +
         \e_a \e_b {\cal H}_{ab}\left(P_a,P_b,P_X\right),
\end{equation}
where
\begin{eqnarray}
     P_a &=& \frac{i}{q^2 - m^2_{\eta_a} +i\e},\ 
     P_b = \frac{i}{q^2 - m^2_{\eta_b} +i\e},\ 
     P_X = \frac{i}{q^2 - m^2_X +i\e}, \,
\nonumber\\
\nonumber\\
\nonumber\\
     {\cal H}_{ab}\left(A,B,C\right) &=& 
           -\frac{1}{3}\left[
             \frac{\left( m^2_{jj}-m^2_{\eta_a}\right)
                   \left( m^2_{rr}-m^2_{\eta_a}\right)}
                  {\left( m^2_{\eta_a} - m^2_{\eta_b}\right)
                   \left( m^2_{\eta_a} - m^2_X\right)}
                 A
            -\frac{\left( m^2_{jj}-m^2_{\eta_b}\right)
                   \left( m^2_{rr}-m^2_{\eta_b}\right)}
                  {\left( m^2_{\eta_a} - m^2_{\eta_b}\right)
                   \left( m^2_{\eta_b} - m^2_X\right)}
                 B \right.\, 
\nonumber\\
&&\qquad\quad\left.
            +\frac{\left( m^2_X-m^2_{jj}\right)
                   \left( m^2_X-m^2_{rr}\right)}
                  {\left( m^2_X-m^2_{\eta_a}\right)
                   \left( m^2_X-m^2_{\eta_b}\right)}
                 C\ \right].
\label{eq:Hfunction}
\end{eqnarray}

\subsection{Baryons containing heavy quarks in \PQCPT} \label{hqbpqcpt}

Heavy quark effective theory (HQET) and \PQCPT\ can be combined to describe
the interactions of heavy hadrons with the pseudo-Goldstone modes. The hadrons
we treat below are baryons composed of one heavy quark $Q$, where $Q = c$ or $b$,
and two light quarks. Let $m_Q$ denote the mass of the heavy quark. In the limit
$m_Q \to \infty$, these baryons (and indeed all heavy hadrons) can be classified by 
the spin of their light degrees of freedom, $s_\ell$.  This classification is possible 
because the heavy quark's spin decouples from the system in the $m_Q \to \infty$ 
limit.  The inclusion of baryons containing a heavy quark into \CPT\ was carried out 
in~\cite{Yan:1992gz,Cho:1992gg,Cho:1992cf} and their masses were calculated to one-loop 
order in~\cite{Savage:1995dw}. The quenched chiral theory for baryons with a heavy quark
was written down in~\cite{Chiladze:1997uq} and the partially quenched theory 
was investigated in~\cite{Arndt:2003vx} for $SU(4|2)$.  Here we include the baryons in 
$SU(6|3)$ \PQCPT.

Consider first the baryons with $s_\ell = 0$. These have spin $\frac{1}{2}$
because there is only one way to combine the spin of the heavy quark with that of the light
degrees of freedom. To include these baryons into \PQCPT, we use the method of interpolating
fields to classify their representations of 
$SU(6|3)$~\cite{Labrenz:1996jy,Savage:2001dy,Chen:2001yi,Beane:2002vq,Arndt:2003vx}. 
The $s_\ell = 0$ baryons are described by the field\footnote{%
Since we shall work only to leading order in $1/m_Q$, we omit the label $Q$ from 
all baryon tensors.
}
\begin{equation}
\cT^\gamma_{ij} 
\sim 
Q^{\gamma,c} 
\left( 
q_i^{\a,a} q_j^{\b,b}
+ 
q_i^{\b,b} q_j^{\a,a} 
\right)
\varepsilon_{abc} (C\gamma_5)_{\a\b}
,\label{eq:Tinterp}
\end{equation}
where $i$ and $j$ are flavor indices. The flavor tensor $\cT$ 
has the property 
\begin{equation}
\cT_{ij} = (-)^{\eta_i \eta_j} \cT_{ji},
\end{equation}
and thus forms a $\mathbf{39}$-dimensional representation of $SU(6|3)$. It is convenient
to classify these states under the quark sectors (valence, sea or ghost) acted upon
and to this end we use the super-algebra terminology of~\cite{Hurni:1981ki,Chen:2001yi}. 
Under $SU(3)_{\text{val}} \otimes SU(3)_{\text{sea}} \otimes SU(3)_{\text{ghost}}$, 
the ground floor, level A transforms as a $(\mathbf{3},\mathbf{1},\mathbf{1})$ and 
contains the familiar $s_\ell = 0$ baryon tensor of QCD $T_{ij}$, explicitly 
$\cT_{ij} = T_{ij}$, when the indices are restricted to the range $1-3$. 
In our normalization convention, we have 
\begin{equation}
T_{ij} 
= 
\frac{1}{\sqrt{2}}
\begin{pmatrix}
0              &     \L_Q      & \Xi_Q^{+\frac{1}{2}} \\
- \L_Q         &        0      & \Xi_Q^{-\frac{1}{2}} \\
- \Xi_Q^{+\frac{1}{2}} & -\Xi_Q^{-\frac{1}{2}} &      0     
\end{pmatrix}_{ij}
\label{eq:TSU3},\end{equation}
where the superscript labels the $3$-projection of isospin. 
The first floor of level A contains nine baryons that transform 
as a $(\mathbf{3},\mathbf{1},\mathbf{3})$, while the ground floor of
level B contains nine baryons that transform as a 
$(\mathbf{3},\mathbf{3},\mathbf{1})$. The remaining floors and levels
are simple to construct but are not necessary for our calculation.

Next we consider the $s_\ell = 1$ baryons. Such baryons come 
in two varieties, spin $\frac{1}{2}$ and $\frac{3}{2}$, because there
are two ways to combine $s_\ell$ with the heavy quark's spin. In the $m_Q \to \infty$
limit, these two multiplets are degenerate and can be described by 
one field $\cS^\mu_{ij}$.  When establishing the representations, we are of course 
ignoring this two-fold degeneracy (which is only lifted away from the heavy quark limit). 
The interpolating field for these baryons is
\begin{equation}
\cS^{\mu, \gamma}_{ij} 
\sim 
Q^{\gamma,c} 
\left( 
q_i^{\a,a} q_j^{\b,b}
- 
q_i^{\b,b} q_j^{\a,a} 
\right)
\varepsilon_{abc} (C\gamma^\mu)_{\a\b}
,\label{eq:Sinterp} 
\end{equation}
where the flavor tensor satisfies
\begin{equation}
\cS^\mu_{ij} = (-)^{1 + \eta_i \eta_j } \cS^\mu_{ji}
,\end{equation}
and hence makes up a $\mathbf{42}$-dimensional representation of $SU(6|3)$. 
Under $SU(3)_{\text{val}} \otimes SU(3)_{\text{sea}} \otimes SU(3)_{\text{ghost}}$, 
the ground floor, level A transforms as a $(\mathbf{6},\mathbf{1},\mathbf{1})$ and 
contains the familiar $s_\ell = 1$ baryon tensor of QCD $S^\mu_{ij}$, explicitly 
$\cS^\mu_{ij} = S^\mu_{ij}$, when the indices are restricted to the range $1-3$.
Here the QCD flavor tensor is given as
\begin{equation}
S^\mu_{ij} = \frac{1}{\sqrt{3}} (v^\mu + \gamma^\mu) \gamma_5 \, B_{ij}  + B^{* \mu}_{ij}
,\label{eq:Sdecomp}
\end{equation} 
with
\begin{equation}
B_{ij} 
= 
\begin{pmatrix}
\Sigma_Q^{+1}                  &  \frac{1}{\sqrt{2}} \Sigma_Q^0   & \frac{1}{\sqrt{2}} \Xi_Q^{\prime + \frac{1}{2}} \\
\frac{1}{\sqrt{2}} \Sigma_Q^0  &  \Sigma_Q^{-1}                   & \frac{1}{\sqrt{2}} \Xi_Q^{\prime - \frac{1}{2}} \\
\frac{1}{\sqrt{2}} \Xi_Q^{\prime + \frac{1}{2}}           &  \frac{1}{\sqrt{2}} \Xi_Q^{\prime - \frac{1}{2}}            & \Omega_Q
\end{pmatrix}_{ij}
\label{eq:BSU3},\end{equation}
and similarly for the $B^{*\mu}_{ij}$. Above $v^\mu$ is the velocity vector of the heavy hadron 
and we have suppressed velocity labels on all heavy hadron fields throughout. 
Again the superscript on these states labels the $3$-projection of isospin. 
The first floor of level A contains nine baryons that transform 
as a $(\mathbf{3},\mathbf{1},\mathbf{3})$, while the ground floor of
level B contains nine baryons that transform as a 
$(\mathbf{3},\mathbf{3},\mathbf{1})$. The remaining floors and levels
are simple to construct but are not necessary for our calculation.

The free Lagrangian for the $\cT$ and $\cS^\mu$ fields is given by
\begin{eqnarray}
\mathfrak{L} 
&=& -
i \left( \ol \cS {}^\mu v \cdot D \cS_\mu \right)
+ 
\D \left( \ol \cS {}^\mu \cS_\mu \right)
+ 
\l_1 \left( \cS {}^\mu  \cM \cS_\mu \right)
+ 
\l_2 \left( \cS {}^\mu \cS_\mu \right) \str \cM
\notag \\
&& + 
i \left( \ol \cT v \cdot D \cT \right) 
+ 
\l_3 \left( \ol \cT \cM \cT \right)
+ 
\l_4 \left( \ol \cT \cT \right) \str \cM 
\label{eq:STfree}
.\end{eqnarray}
Above we have employed $()$-notation for the flavor contractions that are invariant under
chiral transformations. The relevant contractions can be found in~\cite{Arndt:2003vx}.
The mass of the $\cT$ field has been absorbed into the static phase of the heavy hadron fields. 
Thus the leading-order mass splitting $\D$ appears as the mass of the $\cS^\mu$. This splitting remains finite as 
$m_Q \to \infty$ and cannot be removed by field redefinitions due to the interaction of the 
$\cT$ and $\cS^\mu$ fields. In our power counting, we treat $\D \sim m_\pi$. 
The Lagrangian contains the chiral-covariant derivative $D^\mu$, 
the action of which is identical on $\cT$ and $\cS^\mu$ fields and has the form~\cite{Arndt:2003vx}
\begin{equation}
\left( D^\mu \cT \right)_{ij}
=
\partial^\mu \cT_{ij} 
+ 
V^\mu_{ii'} \cT_{i'j} 
+ 
(-)^{\eta_i (\eta_j + \eta_{j'})}
V^\mu_{jj'} \cT_{ij'}
.\end{equation}
The vector and axial-vector meson fields are defined by
\begin{equation}
V^\mu = \frac{1}{2} \left( \xi \partial^\mu \xi^\dagger + \xi^\dagger \partial^\mu \xi \right), 
\qquad 
A^\mu = \frac{i}{2} \left( \xi \partial^\mu \xi^\dagger - \xi^\dagger \partial^\mu \xi \right)
,\end{equation} 
respectively. The mass operator appears as
\begin{equation}
\cM = \frac{1}{2} \left( \xi m_q \xi + \xi^\dagger m_q \xi^\dagger \right)
.\end{equation}
To calculate baryon masses to NLO, the above Lagrangian is not sufficient at tree level. This 
is because the small parameter $\D$ transforms as a singlet under chiral rotations. 
Consequently arbitrary polynomial functions of $\D / \L_\chi$ must multiply terms in the 
Lagrangian, for example one such term is 
\begin{equation}
\mathfrak{L} = 
\l_1^{(1)} 
\frac{\D}{\L_\chi} 
\left( \cS {}^\mu  \cM \cS_\mu \right).
\notag 
\end{equation}
We will not write down all operators that lead to these effects nor will we 
keep track of similar such terms that arise from loop integrals.
To account for such terms to the order we are working, 
we must treat the LECs as linear functions of $\D$. Specifically
\begin{equation}
\l_j \to
\l_j \left( \frac{\D}{\L_\chi} \right)
= 
\l_j^{(0)} + \l_j^{(1)} \frac{\D}{\L_\chi} 
.\end{equation} 
Determination of the LECs for the four omitted $\D/\L_\chi$ operators 
requires the ability to vary $\D$, and for this reason we keep the effect of these operators only implicitly.\footnote{%
One must also consider operators that change the value of the mass splitting in the chiral limit.  
Let us denote $\Delta^{(0)}$ as the mass of the $\cS^\mu$ field in Eq.~\eqref{eq:STfree}. To the order we are working,
we need the following operators
\begin{equation}
\mathfrak{L} = 
\Delta^{(0)} 
\left[ 
\left( \ol \cS {}^\mu \cS_\mu \right) 
+ 
\Delta^{(1)} \frac{\Delta^{(0)}}{\L_\chi}
\left( \ol \cS {}^\mu \cS_\mu \right)
+ 
\Delta^{(2)} 
\left(
\frac{\Delta^{(0)}}{\L_\chi}
\right)^2
\left( \ol \cS {}^\mu \cS_\mu \right)
\right]
\notag
.\end{equation}
An overall factor of $\D^{(0)}$ appears to render the new LECs $\D^{(1)}$ and $\D^{(2)}$ dimensionless. 
The mass splitting $\D$ that results from these contributions is
\begin{equation}
\D = 
\D^{(0)} 
\left[ 
1
+ 
\D^{(1)} \frac{\D^{(0)}}{\L_\chi}
+ 
\D^{(2)} 
\left( 
\frac{\D^{(0)}}{\L_\chi}
\right)^2
\right]
\notag
.\end{equation}
Additionally the loop integrals contribute to $\D$ up to $\cO(\D^{(0)^3})$. Instead of treating
$\D$ as a function of $\D^{(0)}$, we combine the new LECs and contributions to $\D$ from loop integrals
into the parameter $\D$ that appears in Eq.~\eqref{eq:STfree}. Treating $\D$ as the mass of the $\cS^\mu$ field
instead of $\D^{(0)}$ results in differences that are of higher order than we need consider.  
}

The Lagrangian that describes the interactions of the $\cT$ and $\cS^\mu$ fields with 
the pseudo-Goldstone modes is given at leading order by the Lagrangian
\begin{equation}
\mathfrak{L} 
= 
i g_2 
\left( 
\ol \cS {}^\mu v^\nu A^\rho \cS^\sigma 
\right)
\varepsilon_{\mu \nu \rho \sigma}
+
\sqrt{2} \, g_3 
\left[ 
\left( 
\ol \cT A^\mu \cS_\mu 
\right) 
+ 
\left(
\ol \cS {}^\mu A_\mu \cT
\right)
\right],
\label{eq:STM}
\end{equation}
and gives rise to the loop corrections to the baryon masses at NLO. 
The LECs appearing above, the $\l_j$, $g_2$, and $g_3$, all have 
the same numerical values as those used in $SU(3)$ heavy hadron \CPT, 
see Appendix \ref{cpt}.

\section{Masses of the $s_{\ell} = 0$ baryons} \label{triplet}

The mass of the $i^{\text{th}}$ $s_\ell = 0$ baryon, $T_i$,  can be written in the form
\begin{eqnarray}
     M_{T_i} = M_0 \left(\mu \right) -  M_{T_i}^{(1)}\left(\mu \right)
                - M_{T_i}^{(3/2)}\left(\mu \right)
                - \ldots
\label{eq:Tmassexp}
\,.\end{eqnarray}
Here, $M_0 \left(\mu \right)$ is the renormalized mass of the $T$ baryons
in the chiral limit. This quantity is independent of $m_q$ and also of the $T_i$.
$M_{T_i}^{(n)}$ is the contribution to the $i^{th}$ $s_\ell = 0$ baryon mass of order $m_q^{(n)}$, 
and $\mu$ is the renormalization scale. In this calculation we use dimensional regularization
with a modified minimal subtraction scheme, where we have subtracted terms proportional to 
\begin{equation}
\frac{1}{\varepsilon} - \gamma_E + 1 + \log 4 \pi
.\end{equation}

\begin{figure}[tb]
  \centering
  \includegraphics[width=0.5\textwidth]{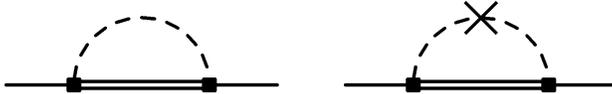}%
  \caption{
    One-loop graphs which give contributions to the masses of the $s_\ell = 0$ baryons.  
    The single, double and dashed lines correspond to  $s_\ell = 0$ baryons,
    $s_\ell = 1$ baryons and mesons, respectively.  The filled squares
    denote the axial coupling given in Eq.~(\ref{eq:STM}), while the cross
    denotes the hairpin interaction. 
  }
  \label{fig:NLOT}
\end{figure}

To calculate the masses of the $s_\ell = 0$ baryons in \PQCPT, 
we need the tree-level contributions from the operators in Eq.~\eqref{eq:STfree}.
These give rise to the leading-order contributions to the mass, namely
\begin{equation}
M_T^{(1)} = \l_3 \,  m_T + \l_4 \, \str (m_q)
,\end{equation}
where the coefficients $m_T$ are listed for individual $T$-states in Table~\ref{t:mT}.

\begin{table}
\caption{The tree-level coefficients in \CPT\ and \PQCPT. The $SU(3)$ and $SU(6|3)$ 
coefficients $m_T$ are listed for the $s_\ell =0$ baryon states $T$.}
%\begin{ruledtabular}
\begin{tabular}{l | c }
$T \qquad$	     & $\qquad m_T \qquad$  \\
\hline
$\L_Q$       
             &  $m_u$ \\
$\Xi_Q$ 
 	     &  $\frac{1}{2} ( m_u + m_s)$ 
\end{tabular}
%\end{ruledtabular}
\label{t:mT}
\end{table}

There are also loop diagrams which are generated from the vertices present in 
Eq.~\eqref{eq:STM}. The relevant loop contributions are depicted 
in Fig.~\ref{fig:NLOT} and include the hairpin interaction from the flavor-neutral field. 
These two one-loop graphs contribute to the $T$ masses at NLO, explicitly we find
\begin{equation}
M_T^{(3/2)} 
=  
\frac{g_3^2}{4 \pi^2 f^2} 
\left[
\sum_\phi 
A_\phi^T \cF(m_\phi, \D, \mu)  
+ 
\sum_{\phi \phi'}
A_{\phi \phi'}^T \cF(m_\phi, m_{\phi'}, \D, \mu)
\right]
\label{eq:MTPQNLO}.\end{equation}
The non-analytic functions $\cF$ appearing in the above equation are defined by
\begin{eqnarray}
\cF (m,\d,\mu) &=& (m^2 - \d^2) 
\left[ \sqrt{\d^2 - m^2}
\log 
\left( 
\frac{\d - \sqrt{\d^2 - m^2 + i \varepsilon}}
{\d + \sqrt{\d^2 - m^2 + i \varepsilon}} \right)
- 
\d \log \left( \frac{m^2}{\mu^2} \right)
\right] \notag \\
&& \phantom{sp} - \frac{1}{2} \d m^2 \log \left( \frac{m^2}{\mu^2} \right)
\label{eq:F}
,\end{eqnarray}
and  
\begin{equation}
\cF (m_\phi, m_{\phi'}, \d, \mu) 
=  
\cH_{\phi \phi'}[\cF(m_\phi,\d,\mu),\cF(m_{\phi'},\d,\mu), \cF(m_X,\d,\mu)]
\label{eq:FPQ}
,\end{equation}
for the hairpin contributions. The first sum in Eq.~\eqref{eq:MTPQNLO}
is over all loop mesons $\phi$ of mass $m_\phi$.  
The second term sums over all pairs of flavor-neutral states in the quark basis, e.g., above  
$\phi\phi'$ runs over $\eta_u \eta_u$, $\eta_u \eta_s$, and $\eta_s \eta_s$; and all other entries are zero. 
In this way there is no double counting. The coefficients $A^T_\phi$, and $A^T_{\phi\phi'}$ appear in 
Table \ref{t:TPQQCD-A} and depend on the particular baryon state $T$.

\begin{table}
\caption{The coefficients $A^T_\phi$ and $A^T_{\phi\phi'}$ in $SU(6|3)$ \PQCPT. Coefficients are
listed for the $s_\ell = 0$ baryon states $T$, and for $A^T_\phi$ are grouped into contributions from loop mesons
with mass $m_\phi$, while for $A^T_{\phi\phi'}$ are grouped into contributions from pairs of quark-basis 
$\eta_q$ mesons.}
%\begin{ruledtabular}
\begin{tabular}{l | c c c c c c c | c c c }
 & \multicolumn{7}{c|}{$A^T_\phi \phantom{ap}$} & \multicolumn{3}{c}{$A^T_{\phi\phi'}$ \phantom{sp}} \\
    & $\quad \pi \quad$ & $\quad K \quad $ & $\quad  \eta_s \quad $ 
    & $ \quad ju \quad$ & $ \quad ru \quad$ 
    & $\quad js \quad$  & $\quad rs \quad$ 
    & $\quad \eta_u \eta_u \quad $ & $\quad \eta_u \eta_s\quad $   & $\quad \eta_s \eta_s \quad$ \\
\hline
$\L_Q$     &  $\frac{1}{2}$ & $0$  & $0$  
           &  $1$ & $\frac{1}{2}$  
           &  $0$ & $0$
           &  $0$ & $0$ & $0$ \\

$\Xi_Q$     
	   &  $0$ & $\frac{1}{2}$  & $0$  
           &  $\frac{1}{2}$ & $\frac{1}{4}$  
           &  $\frac{1}{2}$ & $\frac{1}{4}$
           &  $\frac{1}{4}$ & $-\frac{1}{2}$ & $\frac{1}{4}$ 
\end{tabular}
%\end{ruledtabular}
\label{t:TPQQCD-A}
\end{table}

\section{Masses of the $s_{\ell} = 1$ baryons} \label{six}

The mass of the $i^{\text{th}}$ $s_\ell = 1$ baryon, $B_i$,  can be written in the form
\begin{eqnarray}
     M_{B_i} = M_0 \left(\mu \right) +  \D(\mu) + M_{B_i}^{(1)}\left(\mu \right)
                + M_{B_i}^{(3/2)}\left(\mu \right)
                + \ldots
\label{eq:Bmassexp}
\,.\end{eqnarray}
Here, $M_0 \left(\mu \right)$ is the renormalized mass of the $T$ baryons in the chiral limit
and $\D(\mu)$ is the renormalized splitting between the $T$ and $B$ baryons in the chiral limit. 
Both of these quantities are independent of $m_q$ and also of the $B_i$.
$M_{B_i}^{(n)}$ is the contribution to the $i^{th}$ $s_\ell = 1$ baryon mass of order $m_q^{(n)}$, 
and $\mu$ is the renormalization scale.

To calculate the masses of the $s_\ell = 1$ baryons in \PQCPT, 
we need the tree-level contributions from the operators in Eq.~\eqref{eq:STfree}.
These give rise to the leading-order contributions to the masses, namely
\begin{equation}
M_B^{(1)} = \l_1 \,  m_B + \l_2 \, \str (m_q)
,\end{equation}
where the coefficients $m_B$ are listed for individual $B$-states in Table~\ref{t:mB}.

\begin{table}
\caption{The tree-level coefficients in \CPT\ and \PQCPT. The $SU(3)$ and $SU(6|3)$ 
coefficients $m_B$ are listed for the $s_\ell =1$ baryon states $B$.}
%\begin{ruledtabular}
\begin{tabular}{l | c }
$B \qquad$	     & $\qquad m_B \qquad$  \\
\hline
$\Sigma_Q$       
             &  $m_u$ \\
$\Xi_Q^{\prime}$ 
 	     &  $\frac{1}{2} ( m_u + m_s)$ \\
$\Omega_Q$   
	     & $ m_s$ 
\end{tabular}
%\end{ruledtabular}
\label{t:mB}
\end{table}

\begin{figure}[tb]
  \centering
  \includegraphics[width=0.5\textwidth]{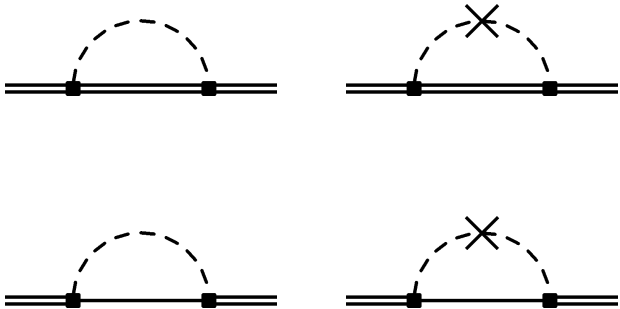}%
  \caption{
    One-loop graphs which give contributions to the masses of the $s_\ell = 1$ baryons.  
    The single, double and dashed lines correspond to  $s_\ell = 0$ baryons,
    $s_\ell = 1$ baryons and mesons, respectively.  The filled squares
    denote the relevant axial couplings given in Eq.~(\ref{eq:STM}), while the crosses
    denote hairpin interactions. 
  }
  \label{fig:NLOS}
\end{figure}

There are also loop diagrams that contribute to the masses. These are generated from the vertices present 
in Eq.~\eqref{eq:STM}. The relevant loop contributions are depicted 
in Fig.~\ref{fig:NLOS} and include hairpin interactions from the flavor-neutral field. 
These four one-loop graphs contribute to the $B$ masses at NLO, explicitly we find
\begin{eqnarray}
M_B^{(3/2)} 
&=&
- \frac{g_3^2}{12 \pi^2 f^2} 
\left[
\sum_\phi 
A_\phi^B \cF(m_\phi, -\D, \mu)  
+ 
\sum_{\phi \phi'}
A_{\phi \phi'}^B \cF(m_\phi, m_{\phi'}, -\D, \mu)
\right]
\notag \\
\label{eq:MBPQNLO}
&& +  
\frac{g_2^2}{12 \pi f^2}
\left[
\sum_\phi  
B_\phi^B m_\phi^3
+ 
\sum_{\phi \phi'}
B_{\phi \phi'}^B \mathcal{M}^3(m_\phi,m_{\phi'})
\right]
.\end{eqnarray}
The non-analytic functions $\cF$ appear in Eqs.~\eqref{eq:F} and \eqref{eq:FPQ}. 
The function $\mathcal{M}$ arising from hairpin contributions is given by
\begin{equation}
\mathcal{M}^3(m_\phi, m_{\phi'}) = \cH_{\phi \phi'} (m_\phi^3, m_{\phi'}^3, m_X^3)
.\end{equation}
The coefficients $A^B_\phi$, and $A^B_{\phi\phi'}$ appear in 
Table \ref{t:BPQQCD-A}, and depend on the particular baryon state $B$. 
The values of $B^B_\phi$ and $B^B_{\phi \phi'}$ are similarly listed in Table~\ref{t:BPQQCD-B}.

\begin{table}
\caption{The coefficients $A^B_\phi$ and $A^B_{\phi\phi'}$ in $SU(6|3)$ \PQCPT. Coefficients are
listed for the $s_\ell = 1$ baryon states $B$, and for $A^B_\phi$ are grouped into contributions from loop mesons
with mass $m_\phi$, while for $A^B_{\phi\phi'}$ are grouped into contributions from pairs of quark-basis 
$\eta_q$ mesons.}
%\begin{ruledtabular}
\begin{tabular}{l | c c c c c c c | c c c }
 & \multicolumn{7}{c|}{$A^B_\phi \phantom{ap}$} & \multicolumn{3}{c}{$A^B_{\phi\phi'}$ \phantom{sp}} \\
    & $\quad \pi \quad$ & $\quad K \quad $ & $\quad  \eta_s \quad $ 
    & $ \quad ju \quad$ & $ \quad ru \quad$ 
    & $\quad js \quad$  & $\quad rs \quad$ 
    & $\quad \eta_u \eta_u \quad $ & $\quad \eta_u \eta_s\quad $   & $\quad \eta_s \eta_s \quad$ \\
\hline
$\Sigma_Q$     
	   &  $-\frac{1}{2}$ & $0$  & $0$  
           &  $1$ & $\frac{1}{2}$  
           &  $0$ & $0$
           &  $0$ & $0$ & $0$ \\

$\Xi^\prime_Q$     
	   &  $0$ & $-\frac{1}{2}$  & $0$  
           &  $\frac{1}{2}$ & $\frac{1}{4}$  
           &  $\frac{1}{2}$ & $\frac{1}{4}$
           &  $\frac{1}{4}$ & $-\frac{1}{2}$ & $\frac{1}{4}$ \\

$\Omega_Q$     
	   &  $0$ & $0$  & $-\frac{1}{2}$  
           &  $0$ & $0$  
           &  $1$ & $\frac{1}{2}$
           &  $0$ & $0$ & $0$ 
\end{tabular}
%\end{ruledtabular}
\label{t:BPQQCD-A}
\end{table}

\begin{table}
\caption{The coefficients $B^B_\phi$ and $B^B_{\phi\phi'}$ in $SU(6|3)$ \PQCPT. Coefficients are
listed for the $s_\ell = 1$ baryon states $B$, and for $B^B_\phi$ are grouped into contributions from loop mesons
with mass $m_\phi$, while for $B^B_{\phi\phi'}$ are grouped into contributions from pairs of quark-basis 
$\eta_q$ mesons.}
%\begin{ruledtabular}
\begin{tabular}{l | c c c c c c c | c c c }
 & \multicolumn{7}{c|}{$B^B_\phi \phantom{ap}$} & \multicolumn{3}{c}{$B^B_{\phi\phi'}$ \phantom{sp}} \\
    & $\quad \pi \quad$ & $\quad K \quad $ & $\quad  \eta_s \quad $ 
    & $ \quad ju \quad$ & $ \quad ru \quad$ 
    & $\quad js \quad$  & $\quad rs \quad$ 
    & $\quad \eta_u \eta_u \quad $ & $\quad \eta_u \eta_s\quad $   & $\quad \eta_s \eta_s \quad$ \\
\hline
$\Sigma_Q$     
	   &  $\frac{1}{2}$ & $0$  & $0$  
           &  $1$ & $\frac{1}{2}$  
           &  $0$ & $0$
           &  $1$ & $0$ & $0$ \\

$\Xi^\prime_Q$     
	   &  $0$ & $\frac{1}{2}$  & $0$  
           &  $\frac{1}{2}$ & $\frac{1}{4}$  
           &  $\frac{1}{2}$ & $\frac{1}{4}$
           &  $\frac{1}{4}$ & $\frac{1}{2}$ & $\frac{1}{4}$ \\

$\Omega_Q$     
	   &  $0$ & $0$  & $\frac{1}{2}$  
           &  $0$ & $0$  
           &  $1$ & $\frac{1}{2}$
           &  $0$ & $0$ & $1$ 
\end{tabular}
%\end{ruledtabular}
\label{t:BPQQCD-B}
\end{table}

\section{Summary} \label{summy}

Above we have determined the masses of both the $s_\ell = 0$ and $s_\ell = 1$ baryons
at one-loop order in the chiral expansion. We performed these partially quenched calculations 
in the isospin limit of $SU(6|3)$ and away from the isospin limit in $SU(4|2)$. 
Understanding the light-quark mass dependence of these baryons' masses in \CPT\ and \PQCPT\ 
is necessary to extrapolate QCD and PQQCD lattice data from the valence and sea quark masses used in the numerical
simulations down to their physical values. Because the LECs of \CPT\ are all contained in \PQCPT, 
fitting PQQCD lattice data to \PQCPT\ expressions can yield rigorous predictions for QCD.

The lattice data, moreover, can be used to determine the LECs appearing in the 
chiral theories. At $\cO(m_q^{3/2})$ in the chiral expansion (and in the heavy quark limit), 
there are only a few LECs which enter into the calculation and 
lattice data can be used to fit these universal parameters.
Of particular interest for study on the lattice
are the properties of the $\Omega_Q$ baryon formed from two strange quarks and a heavy quark. 
This state does not decay strongly and current lattice technology can reach realistic values 
for the strange quark mass. As with the $\Omega^-$~\cite{Toussaint:2004cj}, 
the $\Omega_Q$ is thus an ideal testing ground for obtaining and extrapolating lattice data with dynamical quarks. 
The expressions derived above allow one to analyze the light-quark mass dependence of the 
$\Omega_Q$ mass in partially quenched simulations.  Further work is needed for other processes.

\begin{figure}[tb]
  \centering
  \includegraphics[width=0.45\textwidth]{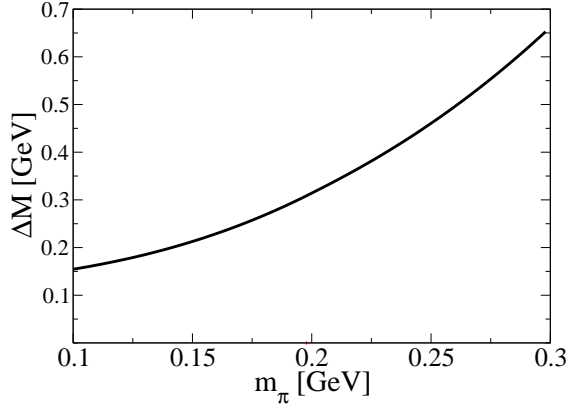}%
  \caption{Pion mass dependence of the splitting $\D M$ between the $s_\ell = 0$ and  $s_\ell = 1$ baryons 
in two flavor \CPT. 
  }
  \label{f:split}
\end{figure}

An example of the utility of \CPT\ is to predict the widths for strong decays of heavy baryons from the lattice, 
e.g.~for $\Sigma_Q \to \L_Q \pi$. Treating the resonant $\Sigma_Q$ state directly on the lattice
is problematic for real-valued Euclidean space simulations. Instead, one can work with 
pion masses $m_\pi \gtrsim 200$ MeV for which the $\Sigma_Q$ is stable, and where one still trusts the chiral expansion.\footnote{%
This is a conservative estimate because finite volume effects and heavy quark effects 
will modify the $\Sigma_Q$ decay threshold. 
}
This enables the lattice simulations to be performed and 
one then uses the \CPT\ or \PQCPT\ expressions derived above to fit the LECs from the lattice data. 
Consequently these LECs lead to predictions for the physical $\Sigma_Q$ resonances. 
For example, their decay widths can be found by knowing the pion mass and the constants $f$, $g_3$, and $\D$. 
The expression for the $\Sigma_Q$ width $\Im \text{m} ( M_{\Sigma_{Q}})$ in terms of these parameters is
\begin{equation}
\Im \text{m} (M_{\Sigma_Q}) 
= 
- \frac{g_{3}^2}{12 \pi f^2} 
(\D^2 - m_\pi^2)^{3/2} .
\end{equation}
As a further application, the mass splitting 
between the $s_\ell = 0$ and $s_\ell = 1$ baryons due to chiral effects can be studied to isolate the various LECs. 
For the real part of the difference between the average $\Sigma_Q$ and $\L_Q$ masses in $SU(2)$ flavor, 
we have
\begin{eqnarray}
\Delta M &\equiv& \Re \text{e} ( \ol M_{\Sigma_Q} - M_{\L_Q}) \notag \\
&=&
\Delta 
+ 
\left( 
\frac{1}{2} \l_1 + \l_2 + \l_4
\right) 
\tr (m_q) 
+ \frac{g_2^2}{12 \pi f^2} m_\pi^3
+ \frac{5 g_3^2}{12 \pi^2 f^2} \cF(m_\pi, \D, \mu) 
.\end{eqnarray}
While the chiral contributions to individual baryon masses are relatively small (na\"{\i}vely they are $\sim 5 \%$), 
the mass splitting between the $s_\ell = 0$ and $s_\ell = 1$ baryons is an order of magnitude more sensitive to 
quark mass variation.\footnote{%
This is also true in \PQCPT, where there is more freedom to vary masses and determine parameters.
}
To see this variation, we plot the mass splitting $\Delta M$  as a function 
of the pion mass in Fig.~\ref{f:split}. In order to plot this splitting, we have assumed the $\l_j$ LECs are of order one at the 
scale $\mu = 1$ GeV, and hence their contribution to the splitting can be neglected. 
We have used the values $g_2 = -1.9$ and $g_3 = 1.5$ as estimates of the couplings. These values come from large $N_c$ 
considerations~\cite{Guralnik:1992dj,Jenkins:1993af}. The parameter $\Delta$ is chosen to be $200$ MeV,  which is near the
phenomenological value.

\begin{figure}[tb]
  \centering
  \includegraphics[width=0.45\textwidth]{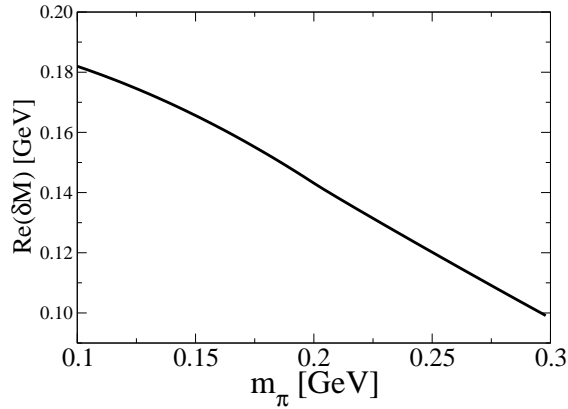}%
  \caption{Mass dependence of the function $\d M$ involving the sextet baryon masses in the isospin limit of \CPT. 
The strange quark mass is fixed using the experimental values for the pion and kaon masses. 
The up quark mass varies and the resulting variation of $\d M$ is plotted as a function of the pion mass. 
  }
  \label{f:rel}
\end{figure}

Considering the sextet baryons in the isospin limit of $SU(3)$, the quark mass dependence from 
chiral loops can be isolated for the linear combination of the masses $\d M$, given by  
\begin{equation}
\delta M = M_{\Sigma_Q} + M_{\Omega_Q} - 2 M_{\Xi_Q^\prime}
.\end{equation}
This is because at LO in the chiral expansion $\d M = 0$, and deviations from zero occur at one-loop order.\footnote{%
Additionally in $SU(6|3)$ \PQCPT, the tree-level contributions to $\d M$ cancel leaving the one-loop results as
the leading non-vanishing contribution. For simplicity we do not consider the extra freedom available  
to extract the couplings by varying the sea quark masses. 
} 
Thus this combination of the masses is only sensitive to the couplings $g_2$ and $g_3$. 
Using our expressions from Appendix~\ref{cpt}, we find
\begin{equation}
\delta M = \frac{1}{12 \pi^2 f^2} \sum_\phi \alpha_\phi \, \left[ \pi \, g_2^2  \, m_\phi^3 + g_3^2 \, \cF(m_\phi, - \D, \mu) \right]  
,\end{equation}
where the coefficient $\a_\phi$ is given by $\a_\pi = 1/ 4$, $\a_K = -1$, and $\a_\eta = 3 / 4$. Notice there is no 
$\mu$-dependence in $\d M$ because of the flavor structure of the coefficient $\a_\phi$, and the Gell-Mann--Oaks--Renner
relation. Using the same estimates of $g_2$, $g_3$ and $\D$ as above, we can plot $\d M$ as a function of the quark masses. 
This is done in Fig.~\ref{f:rel} for the real part of $\d M$. The imaginary part is an order of magnitude smaller. 
For simplicity, we hold the strange quark mass fixed, and vary the up quark mass. 
Results are plotted as a function of the pion mass. 
Determination of the quark mass dependence of $\d M$ provides a clean way to access
the squares of the couplings $g_2$ and $g_3$ from lattice data or partially quenched lattice data.

Soon lattice QCD calculations will enable an understanding of the properties of hadrons 
in terms of quark and gluon degrees of freedom. A crucial first step will be the determination of their masses
from first principles. This requires moving beyond the quenched approximation and including effects from dynamical light quarks. 
Current and foreseeable lattice calculations will rely upon chiral extrapolations
in the light-quark sector. Our expressions for the masses of singly heavy baryons in \CPT\ and \PQCPT\ enable 
such extrapolations. One can go beyond the results here and include higher-order terms in HQET
or try to address lattice artifacts. 
Such a study would improve our ability to extrapolate lattice QCD data on two fronts: 
the light-quark regime and the heavy-quark regime.

\acknowledgments
We thank Tom Mehen for helpful comments. 
This work was supported in part by the U.S. Department of
Energy under Grant No.~DE-FG02-96ER40945.

\appendix

\section{Heavy hadron masses in $SU(4|2)$} \label{pqsutwo}

The set up of heavy hadron chiral perturbation theory in the 
baryon sector for $SU(4|2)$ is quite similar to that of $SU(6|3)$ 
in Sec.~\ref{pqhhcpt}. The partially quenched heavy hadron Lagrangian 
has been written down in~\cite{Arndt:2003vx}. In this appendix, we
briefly review the setup of $SU(4|2)$ PQQCD and \PQCPT. We then 
present the calculation of heavy hadron masses for this graded flavor group.

\subsection{\PQCPT\ for $SU(4|2)$}

In $SU(4|2)$ PQQCD, the Lagrangian in the light-quark sector is
\begin{equation}
\mathfrak{L} = \sum_{j,k=1}^6 \ol{q}_j \left(
  i\Dslash - m_q \right)_{jk} q_k
\label{eq:pqqcdlag2}
,\end{equation}
and differs from the usual two light-flavor Lagrangian of QCD by the
inclusion of four extra quarks; two bosonic ghost quarks, ($\tilde u,
\tilde d$), and two fermionic sea quarks, ($j, l$), in addition
to the light physical quarks ($u, d$).  The six quark fields transform in the 
fundamental representation of the graded $SU(4|2)$ group.  
They appear in the six-component vector
\begin{equation}
q = (u, d, j, l,\tilde{u}, \tilde{d})^{\text{T}}
.\end{equation}
The quark fields obey the same graded equal-time commutation relations as in Eq.~\eqref{eq:qetcr}
but with the grading factor $\eta_k$ now defined by
\begin{equation}
   \eta_k
   = \left\{ 
       \begin{array}{cl}
         1 & \text{for } k=1,2,3,4 \\
         0 & \text{for } k=5,6
       \end{array} 
     \right.
.\end{equation}
The mass matrix with non-degenerate quarks is given by
\begin{equation}
m_q = \diag(m_u, m_d, m_j, m_l, m_u, m_d)
,\end{equation}
where the ghost quarks are still degenerate with their valence partners.
In the limit $m_j \rightarrow m_u$ and $m_l \rightarrow m_d$, 
QCD with two light flavors is recovered.

In the limit of massless light quarks, 
the Lagrangian in Eq. (\ref{eq:pqqcdlag2}) has a graded symmetry 
$SU(4|2)_L \otimes SU(4|2)_R \otimes U(1)_V$, which is
assumed to be spontaneously broken to 
$SU(4|2)_V \otimes U(1)_V$ in analogy with QCD.  
The Lagrangian describing the pseudo-Goldstone mesons of this theory
is identical in form to Eq.~\eqref{eq:pqbosons}. The only difference 
being that the meson fields appearing in Eq.~\eqref{eq:mesonmatrix} are given by 
$\Phi$, which is now an $SU(4|2)$ matrix. 
The upper $2 \times 2$ block of the matrix $M$ contains the familiar
pions, see e.g.~\cite{Beane:2002vq}.
The flavor singlet field that appears in $SU(4|2)$  is defined to be 
$\Phi_0 = {\rm str}( \Phi ) / {\sqrt 2}$.  As before 
the mass of the singlet field $m_0$ can be taken to be
on the order of the chiral symmetry breaking scale.  
In this limit, the $\eta$ two-point correlation functions deviate from their familiar form in \CPT.
For $a,b = u,d,j,l,\tilde u,\tilde d$, the leading-order $\eta_a \eta_b$ propagator is given by
\begin{equation}
{\cal G}_{\eta_a \eta_b} =
        \frac{i \epsilon_a \delta_{ab}}{q^2 - m^2_{\eta_a} +i\epsilon}
        - \frac{i}{2} \frac{\epsilon_a \epsilon_b \left(q^2 - m^2_{jj}
            \right) \left( q^2 - m^2_{ll} \right)}
            {\left(q^2 - m^2_{\eta_a} +i\epsilon \right)
             \left(q^2 - m^2_{\eta_b} +i\epsilon \right)
             \left(q^2 - m^2_X +i\epsilon \right)}\, ,
\end{equation}
where
\begin{equation}
\epsilon_a = (-1)^{1+\eta_a}
.\end{equation}
The mass $m_X$ is now defined as $m_X^2 =
\frac{1}{2}\left(m^2_{jj} + m^2_{ll}\right)$.  
The flavor neutral propagator can be conveniently rewritten as
\begin{equation}
{\cal G}_{\eta_a \eta_b} =
         \e_a \d_{ab} P_a +
         \e_a \e_b {\cal H}_{ab}\left(P_a,P_b,P_X\right),
\end{equation}
where
\begin{eqnarray}
     P_a &=& \frac{i}{q^2 - m^2_{\eta_a} +i\e},\ 
     P_b = \frac{i}{q^2 - m^2_{\eta_b} +i\e},\ 
     P_X = \frac{i}{q^2 - m^2_X +i\e}, \,
\nonumber\\
\nonumber\\
\nonumber\\
     {\cal H}_{ab}\left(A,B,C\right) &=& 
           -\frac{1}{2}\left[
             \frac{\left( m^2_{jj}-m^2_{\eta_a}\right)
                   \left( m^2_{ll}-m^2_{\eta_a}\right)}
                  {\left( m^2_{\eta_a} - m^2_{\eta_b}\right)
                   \left( m^2_{\eta_a} - m^2_X\right)}
                 A
            -\frac{\left( m^2_{jj}-m^2_{\eta_b}\right)
                   \left( m^2_{ll}-m^2_{\eta_b}\right)}
                  {\left( m^2_{\eta_a} - m^2_{\eta_b}\right)
                   \left( m^2_{\eta_b} - m^2_X\right)}
                 B \right.\, 
\nonumber\\
&&\qquad\quad\left.
            +\frac{\left( m^2_X-m^2_{jj}\right)
                   \left( m^2_X-m^2_{ll}\right)}
                  {\left( m^2_X-m^2_{\eta_a}\right)
                   \left( m^2_X-m^2_{\eta_b}\right)}
                 C\ \right].
\label{eq:Hfunction2}
\end{eqnarray}

To include the baryons with one heavy quark into the theory, we use the same interpolating
fields, Eqs.~\eqref{eq:Tinterp} and \eqref{eq:Sinterp}.
The $s_\ell = 0$ baryons are then described by the field $\cT_{ij}$
which forms a $\mathbf{17}$-dimensional representation of $SU(4|2)$. 
The baryon tensor of QCD $T_{ij}$ is contained as $\cT_{ij} = T_{ij}$, 
when the indices are restricted to the range $1-2$. 
In our normalization convention, we have 
\begin{equation}
T_{ij} 
= 
\frac{1}{\sqrt{2}}
\begin{pmatrix}
0              &     \L_Q     \\ 
- \L_Q         &        0     
\end{pmatrix}_{ij}
\label{eq:TSU2}.\end{equation}
The $s_\ell = 1$ baryons are described by $\cS^\mu_{ij}$
which makes up a $\mathbf{19}$-dimensional representation of $SU(4|2)$. 
The baryon tensor of QCD $S^\mu_{ij}$ is embedded as 
$\cS^\mu_{ij} = S^\mu_{ij}$, when the indices are restricted to the range $1-2$.
Here the QCD flavor tensor $S^\mu_{ij}$ is given in terms of $B_{ij}$ and $B^*_{ij}$ as in Eq.~\eqref{eq:Sdecomp}, but with
\begin{equation}
B_{ij} 
= 
\begin{pmatrix}
\Sigma_Q^{+1}                  &  \frac{1}{\sqrt{2}} \Sigma_Q^0  \\ 
\frac{1}{\sqrt{2}} \Sigma_Q^0  &  \Sigma_Q^{-1}                   
\end{pmatrix}_{ij}
\label{eq:BSU2},\end{equation}
and similarly for $B^{*\mu}_{ij}$. The superscript on these states labels the $3$-projection of isospin. 
The remaining states relevant to our calculation have been classified in~\cite{Arndt:2003vx}.

The free Lagrangian for the $\cT$ and $\cS^\mu$ fields is the same as Eq.~\eqref{eq:STfree}
and the Lagrangian that describes the interactions of these fields with 
the pseudo-Goldstone modes is given by Eq.~\eqref{eq:STM}. 
The LECs appearing in the Lagrangian, the $\l_j$, $g_2$, and $g_3$, 
are of course numerically different than those in $SU(6|3)$. 
The constants $\l_1$, $\l_2$, $g_2$ and $g_3$ have the same numerical values as those 
used in $SU(2)$ heavy hadron \CPT. The remaining constants $\l_3^{(PQ)}$ and $\l_4^{(PQ)}$
are different in the partially quenched theory and the LEC $\l_4$ of $SU(2)$ \CPT\
can be found from matching, see Appendix \ref{cpt}.

\subsection{Baryon masses}

For the mass of the $\L_Q$ baryon in $SU(4|2)$ \PQCPT, 
the tree-level operators give the leading-order contributions, namely
\begin{equation}
M_{\L_Q}^{(1)} =  \frac{1}{2} \l_3^{(PQ)} \, \tr (m_q) + \l_4^{(PQ)} \, \str (m_q)
.\end{equation}
The loop diagrams depicted in Fig.~\ref{fig:NLOT} make contributions to the $\L_Q$ mass at NLO
in the chiral expansion. Explicitly we find
\begin{equation}
M_{\L_Q}^{(3/2)} =  
\frac{g_3^2}{4 \pi^2 f^2} 
\left[
\sum_\phi 
A_\phi \cF(m_\phi, \D, \mu)  
+ 
\sum_{\phi \phi'}
A_{\phi \phi'} \cF(m_\phi, m_{\phi'}, \D, \mu)
\right]
.\end{equation}
The non-analytic function $\cF(m,\D,\mu)$ is given in Eq.~\eqref{eq:F}, while
the hairpin loop function $\cF(m_\phi, m_{\phi'},\D,\mu)$ appears in Eq.~\eqref{eq:FPQ}. 
The coefficients $A_\phi$ and $A_{\phi \phi'}$ in the sums over loop mesons and pairs
of flavor-neutral mesons are listed for the $\L_Q$ in Table~\ref{t:TPQQCD2-A}.

\begin{table}
\caption{The coefficients $A_\phi$ and $A_{\phi\phi'}$ in $SU(4|2)$ \PQCPT. Coefficients are
listed for the $\L_Q$, and for $A_\phi$ are grouped into contributions from loop mesons
with mass $m_\phi$, while for $A_{\phi\phi'}$ are grouped into contributions from pairs of quark-basis 
$\eta_q$ mesons.}
%\begin{ruledtabular}
\begin{tabular}{l | c c c c c c c | c c c }
 & \multicolumn{7}{c|}{$A_\phi \phantom{ap}$} & \multicolumn{3}{c}{$A_{\phi\phi'}$ \phantom{sp}} \\
    & $\quad \eta_u \quad$ & $\quad \pi^\pm \quad $ & $\quad  \eta_d \quad $ 
    & $ \quad ju \quad$ & $ \quad lu \quad$ 
    & $\quad jd \quad$  & $\quad ld \quad$ 
    & $\quad \eta_u \eta_u \quad $ & $\quad \eta_u \eta_d\quad $   & $\quad \eta_d \eta_d\quad$ \\
\hline
$\L_Q$     &  $0$ & $\frac{1}{2}$  & $0$  
           &  $\frac{1}{4}$ & $\frac{1}{4}$  
           &  $\frac{1}{4}$ & $\frac{1}{4}$
           &  $\frac{1}{4}$ & $-\frac{1}{2}$ & $\frac{1}{4}$ \\
\end{tabular}
%\end{ruledtabular}
\label{t:TPQQCD2-A}
\end{table}

For the masses of the $s_\ell = 1$ baryons in $SU(4|2)$ \PQCPT, 
we have the tree level contributions
\begin{equation}
M_B^{(1)} = \l_1 \,  m_B + \l_2 \, \str (m_q)
,\end{equation}
where the coefficients $m_B$ for non-degenerate quarks are listed for individual $B$-states in Table~\ref{t:mB2}.

\begin{table}
\caption{The tree-level coefficients in \CPT\ and \PQCPT. The $SU(2)$ and $SU(4|2)$ coefficients $m_B$ are listed for 
the $s_\ell =1$ baryon states $B$.}
%\begin{ruledtabular}
\begin{tabular}{l | c }
$B \qquad$	     & $\qquad m_B \qquad$  \\
\hline
$\Sigma_Q^{+1}$       
             &  $m_u$ \\
$\Sigma_Q^{0}$       
             &  $\frac{1}{2} (m_u + m_d)$ \\
$\Sigma_Q^{-1}$       
             &  $m_d$ 
\end{tabular}
%\end{ruledtabular}
\label{t:mB2}
\end{table}
The loop diagrams that contribute to the mass are depicted 
in Fig.~\ref{fig:NLOS} and these include hairpin interactions. 
At NLO, we find
\begin{eqnarray}
M_B^{(3/2)} 
&=&  
- \frac{g_3^2}{12 \pi^2 f^2} 
\left[
\sum_\phi 
A_\phi^B \cF(m_\phi, -\D, \mu)  
+ 
\sum_{\phi \phi'}
A_{\phi \phi'}^B \cF(m_\phi, m_{\phi'}, -\D, \mu)
\right]
\notag \\
\label{eq:MBPQNLO2}
&& +
\frac{g_2^2}{12 \pi f^2}
\left[
\sum_\phi
B^B_\phi m_\phi^3  
+ 
\sum_{\phi \phi'}
B^B_{\phi \phi'} \mathcal{M}^3 (m_\phi, m_{\phi'})
\right]
.\end{eqnarray}
The non-analytic functions $\cF$ are given in Eqs.~\eqref{eq:F} and \eqref{eq:FPQ}. 
The coefficients $A^B_\phi$ and $A^B_{\phi \phi'}$ in the sums over loop mesons and pairs
of flavor-neutral mesons are listed for the $B$ baryons in Table~\ref{t:BPQQCD2-A}, and
similarly for $B^B_\phi$ and $B^B_{\phi \phi'}$ in Table~\ref{t:BPQQCD2-B}.

\begin{table}
\caption{The coefficients $A^B_\phi$ and $A^B_{\phi\phi'}$ in $SU(4|2)$ \PQCPT. Coefficients are
listed for the $s_\ell = 1$ baryons $B$, and for $A^B_\phi$ are grouped into contributions from loop mesons
with mass $m_\phi$, while for $A^B_{\phi\phi'}$ are grouped into contributions from pairs of quark-basis 
$\eta_q$ mesons.}
%\begin{ruledtabular}
\begin{tabular}{l | c c c c c c c | c c c }
 & \multicolumn{7}{c|}{$A^B_\phi \phantom{ap}$} & \multicolumn{3}{c}{$A^B_{\phi\phi'}$ \phantom{sp}} \\
    & $\quad \eta_u \quad$ & $\quad \pi^\pm \quad $ & $\quad  \eta_d \quad $ 
    & $ \quad ju \quad$ & $ \quad lu \quad$ 
    & $\quad jd \quad$  & $\quad ld \quad$ 
    & $\quad \eta_u \eta_u \quad $ & $\quad \eta_u \eta_d\quad $   & $\quad \eta_d \eta_d\quad$ \\
\hline
$\Sigma_Q^{+1}$  
	   &  $-\frac{1}{2}$ & $0$  & $0$  
           &  $\frac{1}{2}$ & $\frac{1}{2}$  
           &  $0$ & $0$
           &  $0$ & $0$ & $0$ \\

$\Sigma_Q^{0}$  
	   &  $0$ & $-\frac{1}{2}$  & $0$  
           &  $\frac{1}{4}$ & $\frac{1}{4}$  
           &  $\frac{1}{4}$ & $\frac{1}{4}$
           &  $\frac{1}{4}$ & $-\frac{1}{2}$ & $\frac{1}{4}$ \\

$\Sigma_Q^{-1}$  
	   &  $0$ & $0$  & $-\frac{1}{2}$  
           &  $0$ & $0$  
           &  $\frac{1}{2}$ & $\frac{1}{2}$
           &  $0$ & $0$ & $0$ \\
\end{tabular}
%\end{ruledtabular}
\label{t:BPQQCD2-A}
\end{table}

\begin{table}
\caption{The coefficients $B^B_\phi$ and $B^B_{\phi\phi'}$ in $SU(4|2)$ \PQCPT. Coefficients are
listed for the $s_\ell = 1$ baryons $B$, and for $B^B_\phi$ are grouped into contributions from loop mesons
with mass $m_\phi$, while for $B^B_{\phi\phi'}$ are grouped into contributions from pairs of quark-basis 
$\eta_q$ mesons.}
%\begin{ruledtabular}
\begin{tabular}{l | c c c c c c c | c c c }
 & \multicolumn{7}{c|}{$B^B_\phi \phantom{ap}$} & \multicolumn{3}{c}{$B^B_{\phi\phi'}$ \phantom{sp}} \\
    & $\quad \eta_u \quad$ & $\quad \pi^\pm \quad $ & $\quad  \eta_d \quad $ 
    & $ \quad ju \quad$ & $ \quad lu \quad$ 
    & $\quad jd \quad$  & $\quad ld \quad$ 
    & $\quad \eta_u \eta_u \quad $ & $\quad \eta_u \eta_d\quad $   & $\quad \eta_d \eta_d\quad$ \\
\hline
$\Sigma_Q^{+1}$  
	   &  $\frac{1}{2}$ & $0$  & $0$  
           &  $\frac{1}{2}$ & $\frac{1}{2}$  
           &  $0$ & $0$
           &  $1$ & $0$ & $0$ \\

$\Sigma_Q^{0}$  
	   &  $0$ & $\frac{1}{2}$  & $0$  
           &  $\frac{1}{4}$ & $\frac{1}{4}$  
           &  $\frac{1}{4}$ & $\frac{1}{4}$
           &  $\frac{1}{4}$ & $\frac{1}{2}$ & $\frac{1}{4}$ \\

$\Sigma_Q^{-1}$  
	   &  $0$ & $0$  & $\frac{1}{2}$  
           &  $0$ & $0$  
           &  $\frac{1}{2}$ & $\frac{1}{2}$
           &  $0$ & $0$ & $1$ \\
\end{tabular}
%\end{ruledtabular}
\label{t:BPQQCD2-B}
\end{table}

\section{Heavy hadron masses in $SU(3)$ and $SU(2)$} \label{cpt}

For completeness, we include calculations of the baryon masses in ordinary \CPT. 
These were calculated to one-loop order in~\cite{Savage:1995dw}.
To compare with the main text, we consider the case of $SU(3)$ flavor in the isospin limit
and to compare with the results of Appendix~\ref{pqsutwo}, we consider non-degenerate $SU(2)$. 
For these calculations, we retain the tensors $T_{ij}$ and $S^\mu_{ij}$ in \CPT. For the 
case of $SU(3)$ flavor the tensors are given in Eqs.~\eqref{eq:TSU3} and \eqref{eq:BSU3}, respectively; while
for $SU(2)$ flavor, they are given by Eqs.~\eqref{eq:TSU2} and \eqref{eq:BSU2}, respectively.

The free Lagrangian for the $T_{ij}$ and $S^\mu_{ij}$ fields in $SU(3)$ and $SU(2)$ both have the form
\begin{eqnarray}
\mathfrak{L} 
&=& -
i \left( \ol S {}^\mu v \cdot D S_\mu \right)
+ 
\D \left( \ol S {}^\mu S_\mu \right)
+ 
\l_1 \left( \ol S {}^\mu  \cM S_\mu \right)
+ 
\l_2 \left( \ol S {}^\mu S_\mu \right) \tr \cM
\notag \\
&& + 
i \left( \ol T v \cdot D T \right) 
+ 
\l_3 \left( \ol T \cM T \right)
+ 
\l_4 \left( \ol T T \right) \tr \cM 
\label{eq:STfreeCPT}
.\end{eqnarray}
The only exception is that for the $SU(2)$ flavor group, the term with $\l_3$ is
not independent; hence, one can choose $\l_3 = 0$. When matching the $SU(4|2)$ Lagrangian
onto the $SU(2)$ Lagrangian, one finds $\l_4 = \frac{1}{2} \l_3^{(PQ)} + \l_4^{(PQ)}$. 

The interaction Lagrangian for $T_{ij}$ and $S^\mu_{ij}$ has the same form for both $SU(3)$ and $SU(2)$, namely
\begin{equation}
\mathfrak{L} 
= 
i g_2 
\left( 
\ol S {}^\mu v^\nu A^\rho S^\sigma 
\right)
\varepsilon_{\mu \nu \rho \sigma}
+
\sqrt{2} \, g_3 
\left[ 
\left( 
\ol T A^\mu S_\mu 
\right) 
+ 
\left(
\ol S {}^\mu A_\mu T
\right)
\right].
\label{eq:STMCPT}
\end{equation}
The factor $\sqrt{2}$ is chosen so that $g_3$ agrees with the literature, e.g.~\cite{Manohar:2000dt}.
The numerical values of the $\l_j$, $g_2$, and $g_3$ are of course different for the $SU(3)$ and $SU(2)$ theories.

\subsection{$SU(3)$}

The masses of the $s_\ell = 0$ baryons in three flavor \CPT\ 
are given at LO by
\begin{equation}
M_T^{(1)} = \l_3 \,  m_T + \l_4 \, \tr (m_q)
,\end{equation}
where the coefficients $m_T$ are listed for individual $T$-states in Table~\ref{t:mT}. 
At NLO one has contributions from the loop diagrams in Fig.~\ref{fig:NLOT}, 
with the exception that there is no hairpin interaction in \CPT. 
The expression for the mass is thus
\begin{equation}
M_T^{(3/2)} =
\frac{g_3^2}{4 \pi^2 f^2}
\sum_\phi A_\phi^T \cF(m_\phi, \D, \mu)
,\end{equation}
where the function $\cF$ appears in Eq.~\eqref{eq:F} and the coefficients
$A_\phi^T$ are listed in Table~\ref{t:TQCD-A}.

\begin{table}
\caption{The coefficients $A_\phi^T$ in $SU(3)$ \CPT. Coefficients are
listed for the $s_\ell = 0$ baryon states $T$.}
%\begin{ruledtabular}
\begin{tabular}{l | c c c }
    & \multicolumn{3}{c}{$A_\phi^T$} \\
$T$    & $\quad \pi \quad$ & $\quad K \quad$ & $\quad \eta \quad$ \\
\hline
$\L_Q$       
           & $\frac{3}{2}$  &  $\frac{1}{2}$ &  $0$    \\
$\Xi_Q$ 
           & $\frac{3}{8}$ &  $\frac{5}{4}$  &  $\frac{3}{8}$\\
 \end{tabular}
%\end{ruledtabular}
\label{t:TQCD-A}
\end{table}

Next the masses of the $s_\ell =1$ baryons in three flavor \CPT\ 
are given at LO by
\begin{equation}
M_B^{(1)} = \l_1 \,  m_B + \l_2 \, \tr (m_q)
,\end{equation}
where the coefficients $m_B$ for quarks are listed for individual $B$-states in Table~\ref{t:mB}. 
The loop diagrams that contribute to the mass are depicted 
in Fig.~\ref{fig:NLOS}, excluding the diagrams with hairpin interactions. 
The NLO expression for the mass is 
\begin{eqnarray}
M_B^{(3/2)} 
&=&  
- \frac{g_3^2}{12 \pi^2 f^2}
\sum_\phi  
A_\phi^B \cF(m_\phi, -\D, \mu)
+
\frac{g_2^2}{12 \pi f^2}
\sum_\phi 
B_\phi^B m_\phi^3
.\end{eqnarray}
where the function $\cF$ appears in Eq.~\eqref{eq:F}, and the coefficients
$A_\phi^B$ are listed in Table~\ref{t:BQCD-A}, while $B_\phi^B$ appear
in Table~\ref{t:BQCD-B}.

\begin{table}
\caption{The coefficients $A_\phi^B$ in $SU(3)$ \CPT. Coefficients are
listed for the $s_\ell = 1$ baryon states $B$.}
%\begin{ruledtabular}
\begin{tabular}{l | c c c }
    & \multicolumn{3}{c}{$A_\phi^B$} \\
$B$    & $\quad \pi \quad$ & $\quad K \quad$ & $\quad \eta \quad$ \\
\hline
$\Sigma_Q$       
           & $\frac{1}{2}$  &  $\frac{1}{2}$ &  $0$    \\
$\Xi^\prime_Q$ 
           & $\frac{3}{8}$ &  $\frac{1}{4}$  &  $\frac{3}{8}$\\
$\Omega_Q$ 
           & $0$ &  $1$  &  $0$\\
\end{tabular}
%\end{ruledtabular}
\label{t:BQCD-A}
\end{table}

\begin{table}
\caption{The coefficients $B_\phi^B$ in $SU(3)$ \CPT. Coefficients are
listed for the $s_\ell = 1$ baryon states $B$.}
%\begin{ruledtabular}
\begin{tabular}{l | c c c }
    & \multicolumn{3}{c}{$B_\phi^B$} \\
$B$    & $\quad \pi \quad$ & $\quad K \quad$ & $\quad \eta \quad$ \\
\hline
$\Sigma_Q$       
           & $1$  &  $\frac{1}{2}$ &  $\frac{1}{6}$    \\
$\Xi^\prime_Q$ 
           & $\frac{3}{8}$ &  $\frac{5}{4}$  &  $\frac{1}{24}$\\
$\Omega_Q$ 
           & $0$ &  $1$  &  $\frac{2}{3}$\\
\end{tabular}
%\end{ruledtabular}
\label{t:BQCD-B}
\end{table}

\subsection{$SU(2)$}

The mass of the $\L_Q$ baryon in two flavor \CPT\ 
is given at LO by
\begin{equation}
M_{\L_Q}^{(1)} =  \l_4 \, \tr (m_q)
.\end{equation}
The loop diagram without hairpin interaction shown in Fig.~\ref{fig:NLOT} 
makes the NLO contribution which reads 
\begin{equation}
M_{\L_Q}^{(3/2)} = \frac{3 g_3^2}{8 \pi^2 f^2} \cF(m_\pi, \D, \mu)
,\end{equation}
with the function $\cF$ defined in Eq.~\eqref{eq:F}.

Finally the masses of the $s_\ell =1$ baryons in two flavor \CPT\ 
are given at LO by
\begin{equation}
M_B^{(1)} = \l_1 \,  m_B + \l_2 \, \tr (m_q)
,\end{equation}
where the coefficients $m_B$ for non-degenerate quarks are listed for individual $B$-states in Table~\ref{t:mB2}. 
The NLO expression for these masses arises from the loop diagrams of Fig.~\ref{fig:NLOS} excluding the hairpins. 
We find 
\begin{eqnarray}
M_B^{(3/2)} 
&=&  
- \frac{g_3^2}{24 \pi^2 f^2}
\cF(m_\pi, -\D, \mu) 
+ 
\frac{g_2^2}{12 \pi f^2} m_\pi^3
.\end{eqnarray}

\bibliography{hb}
\end{document}